\documentclass[
aps,
prb,
superscriptaddress,
twocolumn,
floatfix]{revtex4-2}
\usepackage{amsmath}
\usepackage{amssymb}
\usepackage{graphicx}% Include figure files
\usepackage{dcolumn}% Align table columns on decimal point
\usepackage{bm}% bold math
\usepackage[mathlines]{lineno}% Enable numbering of text and display math
\usepackage{xcolor}
\usepackage{braket}

\usepackage{times}
\usepackage{hyperref}% add hypertext capabilities
\usepackage{wasysym} % for the diameter command

\newcommand{\gvs}{GaV$_4$S$_8$}

\begin{document}

\preprint{APS/123-QED}

\title{Spin dynamics in the skyrmion-host lacunar spinel GaV$_4$S$_8$}
\thanks{This manuscript has been authored by UT-Battelle, LLC under Contract No. DE-AC05-00OR22725 with the U.S. Department of Energy.  The United States Government retains and the publisher, by accepting the article for publication, acknowledges that the United States Government retains a non-exclusive, paid-up, irrevocable, world-wide license to publish or reproduce the published form of this manuscript, or allow others to do so, for United States Government purposes.  The Department of Energy will provide public access to these results of federally sponsored research in accordance with the DOE Public Access Plan (http://energy.gov/downloads/doe-public-access-plan).}

\author{G. Pokharel}
\email{ganeshpokharel@ucsb.edu}
\affiliation{Department of Physics \& Astronomy, University of Tennessee, Knoxville, TN 37996, USA}
\affiliation{Materials Science \& Technology Division, Oak Ridge National Laboratory, Oak Ridge, TN 37831, USA}
\affiliation{Materials Department and California Nanosystems Institute, University of California Santa Barbara, Santa Barbara, CA 93106, USA}

\author{H. Suriya Arachchige}
\affiliation{Department of Physics \& Astronomy, University of Tennessee, Knoxville, TN 37996, USA}
\affiliation{Materials Science \& Technology Division, Oak Ridge National Laboratory, Oak Ridge, TN 37831, USA}

\author{S. Gao}
\affiliation{Materials Science \& Technology Division, Oak Ridge National Laboratory, Oak Ridge, TN 37831, USA}
\affiliation{Neutron Scattering Division, Oak Ridge National Laboratory, Oak Ridge, TN 37831, USA}

\author{S. -H. Do}
\affiliation{Materials Science \& Technology Division, Oak Ridge National Laboratory, Oak Ridge, TN 37831, USA}

\author{R. S. Fishman}
\affiliation{Materials Science \& Technology Division, Oak Ridge National Laboratory, Oak Ridge, TN 37831, USA}

\author{G. Ehlers}
\affiliation{Neutron Technologies Division, Oak Ridge National Laboratory, Oak Ridge, TN 37831, USA} 

\author{Y. Qiu}
\affiliation{NIST Center for Neutron Research, National Institute of Standards and Technology, Gaithersburg, MD 20899, USA}

\author{J. A. Rodriguez-Rivera}
\affiliation{NIST Center for Neutron Research, National Institute of Standards and Technology, Gaithersburg, MD 20899, USA}

\author{M. B. Stone}
\affiliation{Neutron Scattering Division, Oak Ridge National Laboratory, Oak Ridge, TN 37831, USA}

\author{H. Zhang}
\affiliation{Department of Physics \& Astronomy, University of Tennessee, Knoxville, TN 37996, USA}

\author{S. D. Wilson}
\affiliation{Materials Department and California Nanosystems Institute, University of California Santa Barbara, Santa Barbara, CA 93106, USA}

\author{D. Mandrus}
\affiliation{Department of Materials Science \& Engineering, University of Tennessee, Knoxville, TN 37996, USA}
\affiliation{Materials Science \& Technology Division, Oak Ridge National Laboratory, Oak Ridge, TN 37831, USA}
\affiliation{Department of Physics \& Astronomy, University of Tennessee, Knoxville, TN 37996, USA}

\author{A. D. Christianson}
\email{christiansad@ornl.gov}
\affiliation{Materials Science \& Technology Division, Oak Ridge National Laboratory, Oak Ridge, TN 37831, USA}

\date{\today}

\begin{abstract}
In the lacunar spinel GaV$_4$S$_8$, the interplay of spin, charge, and orbital degrees of freedom produces a rich phase diagram that includes an unusual N\'eel-type skyrmion phase composed of molecular spins. To provide insight into the interactions underlying this complex phase diagram, we study the spin excitations in GaV$_4$S$_8$ through inelastic neutron scattering measurements on polycrystalline and single crystal samples. Using linear spin wave theory, we describe the spin wave excitations using a model where V$_4$ clusters decorate an FCC lattice.  The effective cluster model includes a ferromagnetic interaction and a weaker antisymmetric Dzyaloshinskii-Moriya (DM) interaction between the neighboring molecular spins. Our work clarifies the spin interactions in GaV$_4$S$_8$ and supports the picture of interacting molecular clusters.

\end{abstract}

\pacs{Valid PACS appear here}% PACS, the Physics and Astronomy
\maketitle

\section{Introduction}

% Magnetic skyrmions are topological spin textures characterized by an integer topological number. These quasiparticles are considered to have great potential for applications in spintronics devices \cite{nagaosa_topological_2013, Finocchio_2016, fert_magnetic_2017, back_skyrmionics_2020, ZhangX_2020}. Since the first discovery of the Bloch-type skyrmions in MnSi~\cite{muhlbauer_skyrmion_2009}, different types of magnetic skyrmions have been identified~\cite{nayak_magnetic_2017, yu_transformation_2018, gao_fractional_2020}. For instance, N\'eel-type skrymion spin textures, first realized in the polar system GaV$_4$S$_8$, have attracted much attention~\cite{Loidl1500916,Ruff_2015,Widmann_2017, kurumaji_neel_2017, white_direct_2018, Ellie_2020}. Distinct from the tangential rotation of the spin whorls in Bloch-type skyrmions, N\'eel-type skyrmions are composed of spins rotating in radial planes, which provides additional opportunities for skyrmion manipulations~\cite{woo_observation_2016, woo_current_2018, wu_neel_2020}.

Magnetic skyrmions are topological spin textures characterized by an integer topological number and have potential for applications in spintronics devices \cite{nagaosa_topological_2013, Finocchio_2016, fert_magnetic_2017, back_skyrmionics_2020, ZhangX_2020}. Since the first discovery of the Bloch-type skyrmions in MnSi~\cite{muhlbauer_skyrmion_2009}, different types of magnetic skyrmions have been identified~\cite{nayak_magnetic_2017, yu_transformation_2018, gao_fractional_2020}. Among them, the N\'eel-type skrymion first realized in the polar system GaV$_4$S$_8$ has drawn great attention~\cite{Loidl1500916,Ruff_2015,Widmann_2017, kurumaji_neel_2017, white_direct_2018, Ellie_2020}. Distinct from the spin whorls in the Bloch-type skyrmions, N\'eel-type skyrmions are composed of spins rotating in the radial planes, which provides new opportunities for skyrmion manipulations~\cite{woo_observation_2016, woo_current_2018, wu_neel_2020}.

\begin{figure}[t]
\centering
\includegraphics[width=1\columnwidth]{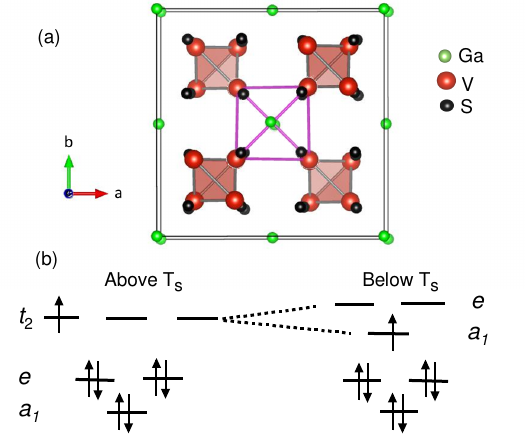}
    \caption{
    (a) Crystal structure of GaV$_4$S$_8$ highlighting the breathing pyrochlore V-sublattice composed of a periodic arrangement of large and small V$_4$ tetrahedra. (b) Molecular orbital diagrams of the V$_4$ tetrahedra in GaV$_4$S$_8$ with (below T$_{\text{s}}$) and without (above T$_{\text{s}}$) the Jahn-Teller distortion~\cite{Xu_Ke_2015}.
    }
\label{str}
\end{figure}

% In this paper we focus on GaV$_4$S$_8$, which has emerged as an important host of the N\'eel-type skyrmion lattice.  
GaV$_4$S$_8$ belongs to the lacunar spinel family of compounds with a chemical formula $AM_4X_8$ ($A$ = Ga, Ge; $M$ = V, Mo, Nb, and Ta; $X$ = S and Se)~\cite{Abd_2004, fujima_thermo_2017, Ruff_2017, Dorolti_2010, Ta_2013, Stoliar_2013, Singh_2014, Geirhos_2018, Heung_2014, Zhang_2019, Matt_2017,Reschke_2020}, where the magnetic $M$ ions form a breathing pyrochlore lattice as shown in Fig.~\ref{str}(a). \gvs{} is a magnetic semiconductor where the magnetism is associated with the unpaired valence electrons localized in the V$_4$ tetrahedra ~\cite{muller_2006}. Assuming Ga and S take their most common valence configuration, charge balance yields Ga$^{3+}$V$_4^{3.25+}$S$_8^{2-}$ with 7 valence electrons per V$_4$ cluster. Due to the short V-V metal bonds in the smaller V$_4$ tetrahedra, the V $3d$-bands hybridize with each other, leading to a molecular spin-1/2 that is almost evenly distributed across the V$_4$ tetrahedra~\cite{Rebecca_2020}. At temperatures below T$_{\text{s}}\sim44$ K, GaV$_4$S$_8$ undergoes a cubic ($F\overline{4}3m$)-rhombohedral ($R3m$) transition due to a Jahn-Teller (JT) distortion, which elongates the V$_4$ tetrahedra along the \{111\} directions and results in an easy axis anisotropy for the $V_4$ spins as shown in Fig.~\ref{str}. This anisotropy stabilizes a ferromagnetic (FM) ground state below $T_{c}=6$~K~\cite{Loidl1500916,Ruff_2015,white_direct_2018}. For temperatures between $T_c$ and the long range order transition of $T_N=12.8$~K, GaV$_4$S$_8$ has a thermal-fluctuation-stabilized cycloidal phase with a periodicity of $\sim20$~nm.  The N\'eel-type skyrmion phase emerges from the cycloidal phase in an applied magnetic field ~\cite{Loidl1500916,Ruff_2015,white_direct_2018}.  

 Although the competition between the ferromagnetic exchange interactions and the Dzyaloshinskii-Moriya (DM) interaction has been proposed to induce the spin winding in GaV$_4$S$_8$, the interactions between the $V_4$ spins are still not determined experimentally. Here we report an inelastic neutron scattering (INS) study of the spin excitations in GaV$_4$S$_8$ and determine the coupling strengths between the $V_4$ spins. The powder averaged INS spectra reveal spin wave excitations with a bandwidth of $\sim6$~meV, and the wave vector dependence of the scattering intensity is strongly modulated by the V$_4$ structure factor. For the  crystal sample, dispersive spin wave excitations are observed along the high symmetry directions, and the spectra are reproduced using a classical Heisenberg model that incorporates the ferromagnetic exchange interactions and the DM interactions.
 % with a strength of $\sim15\%$ of the nearest neighbor (NN) ferromagnetic couplings. 

\begin{figure}
\centering
\includegraphics[width=1\columnwidth]{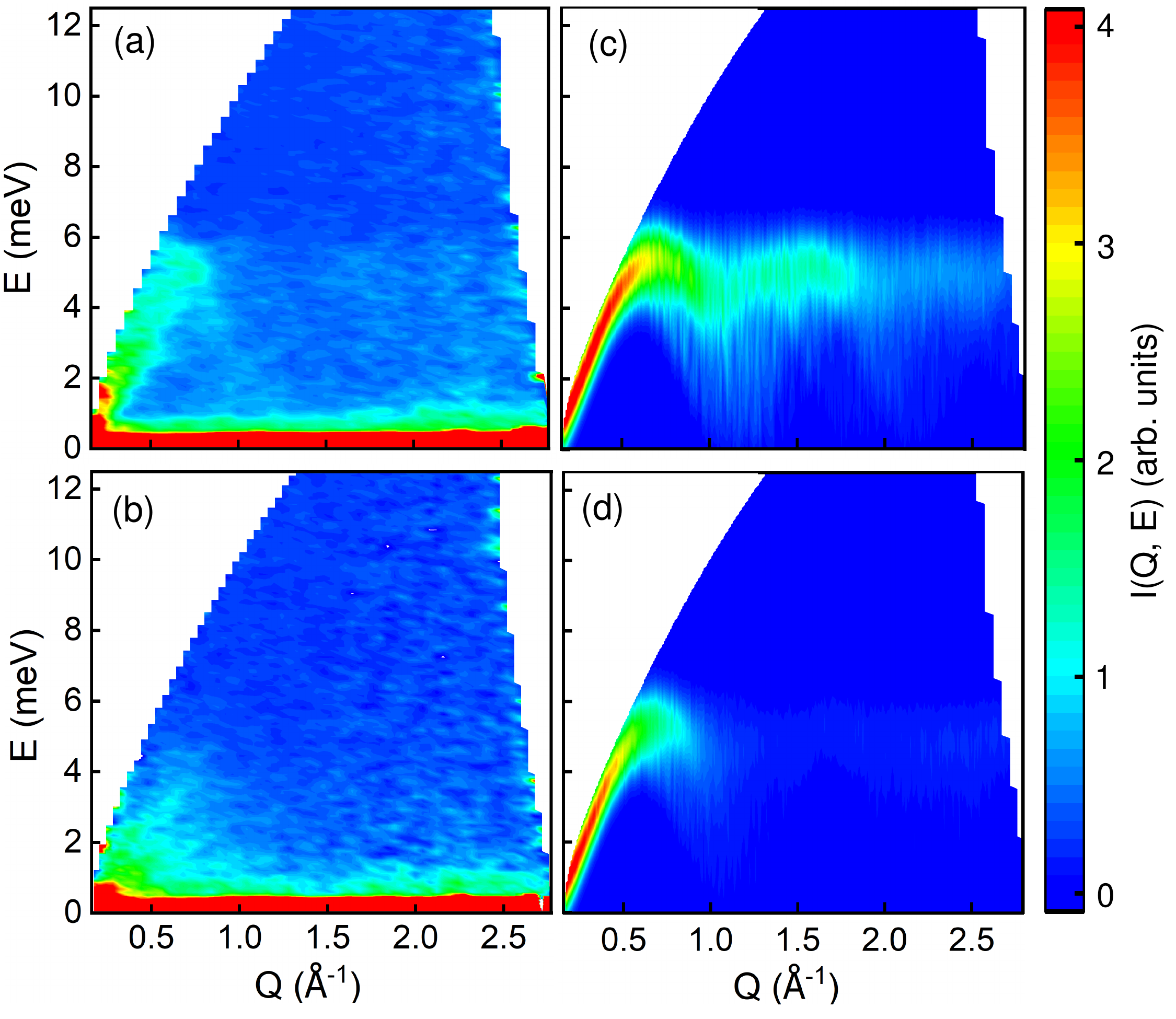}
      \caption{Powder-averaged INS spectra of GaV$_4$S$_8$ measured on SEQUOIA with E$_i$ = 18 meV at (a) 5 K in the FM ordered magnetic phase and (b) 20 K in the paramagnetic phase. (c) and (d) shows the calculated powder-averaged spin excitation spectra in the magnetically ordered state of GaV$_4$S$_8$. The form factor of a single V$^{3+}$ ion is taken into account in  (c), whereas the form factor of  V$_4$ tetrahedra is taken into account in (d) demonstrating the importance of the molecular nature of the V$_4$ tetrahedra. 
}
\label{Powder_INS_GaVS}
\end{figure}

\section{Experimental Details}

Polycrystalline samples of \gvs{} were synthesized by solid state reaction. Stoichiometric amounts of Ga (99.999\%), V (99.5\%), and S (99.9995\%), purchased from Alfa Aesar, were ground together inside a glove box and then pressed into a 0.5-inch diameter pellet. The pellet was slowly heated to 800~$^{\circ}$C over a period of 24 hours and the phase purity was checked by Cu K-alpha x-ray diffraction (XRD). The heating process was repeated until the XRD patterns indicated the presence of a single phase.

\gvs{} crystals were grown by the chemical vapor transport (CVT) technique with iodine as the transport agent.  Approximately 5 weight percent of iodine was mixed with a 2.5 g polycrystalline sample of GaV$_4$S$_8$ and then sealed in a quartz tube. The tube was placed inside a two-zone furnace and a temperature gradient of approximately 50~$^\circ$C was maintained. The temperatures of the hot end and cold end of the quartz tube were regulated to 850$^{\circ}$C and 800$^{\circ}$C respectively. The growth period was about eight weeks. The mass of crystals grown with this protocol and used in the neutron scattering measurements was $\sim100$~mg.

\begin{figure*}[t]
\centering
\includegraphics[width=2\columnwidth]{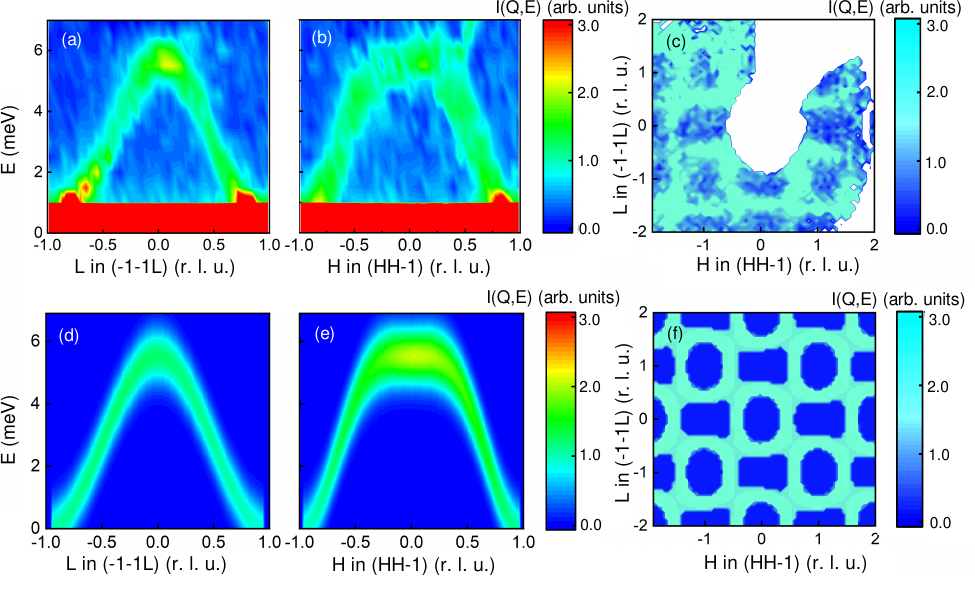}
\caption{Inelastic neutron scattering intensity as a function of energy and wave-vector transfer for GaV$_4$S$_8$ at T=1.7 K. Spin wave excitations along the (a) ($-1-1L$) and (b) ($HH-1$) directions at  $T=1.7$ K  measured using the MACS instrument. (c) Constant-energy slice of the INS spectrum at $E = 3$ meV in the ($HHL$) plane. Calculated spin wave excitation spectra along the (d) ($-1-1L$) (e) ($HH-1$) directions. (f) Calculated constant-energy slice at $E = 3$ meV in the ($HHL$) plane. Measurements and calculations are further described in the text.  
    }
\label{INS_1}
\end{figure*}

The INS measurements of polycrystalline samples were performed on the Fine-Resolution Fermi Chopper Spectrometer SEQUOIA~\cite{Granroth_2010} at the Spallation Neutron Source (SNS) at Oak Ridge National Laboratory (ORNL) with incident energies, $E_i$s of 18 and 40~meV. The sample consisted of a finely ground polycrystalline sample with a mass of 5 g packed inside an aluminum cylinder ($\diameter = 9$~mm) and mounted on a closed-cycle refrigerator (CCR). The measurements were  performed in the high resolution mode with the Fermi chopper set to 300 and 240 Hz for $E_i=40$ and 18 meV, respectively. No magnetic scattering above 7 meV was discernible in the $E_i=40$~meV data and hence the corresponding data is not shown here.  To remove the background contribution, the scattering from an empty sample holder has been subtracted from the data presented in Fig. \ref{Powder_INS_GaVS}.
% The energy resolution of the measurements with E$_i$ = 18 meV is approximately 0.5 meV.

The INS measurements of single crystals were performed using the Multi-Axis Chopper Spectrometer MACS~\cite{Rodriguez_2008} at the NIST center for neutron research (NCNR). Thirteen crystals with a total mass of $\sim 1.2$~g were coaligned  with the ($HHL$) plane horizontal. For the MACS experiments, a helium flow cryostat with a base temperature of 1.7 K was employed. A Be-filter was in placed between the sample and analyser. To cover a large range of momentum  transfer, the sample was rotated about the vertical axis over a range of 120$^{\circ}$ in 2$^{\circ}$ steps. The measurements were performed with a fixed final energy, $E_f$, of 5 meV. A measurement of an empty sample holder was collected with the same configuration for background subtraction. All single crystal INS measurements shown have had this background subtraction applied unless otherwise noted.

\begin{figure} 
\centering
\includegraphics[width=1\columnwidth]{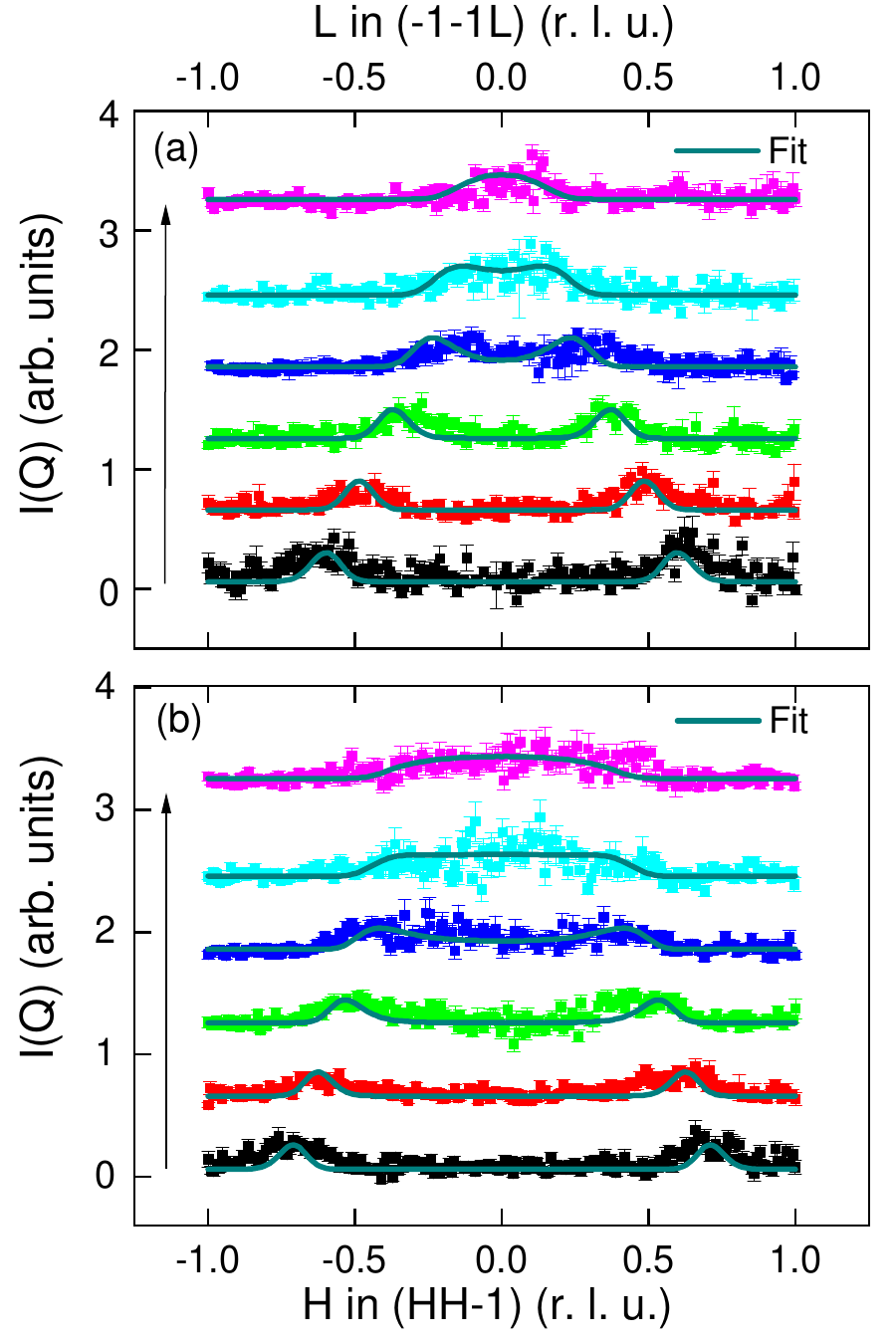}
      \caption{Experimental (data points) and calculated (solid lines) constant-energy cuts at $E$ = 2 meV (black squares), 3 meV (red squares), 4 meV (green squares), 5 meV (blue squares), 5.5 meV (cyan squares) and 6 meV (magneta squares) at 1.7 K (a) along the ($-1-1L$) direction and (b) along the ($HH-1$) direction. Error bars represent one standard deviation. The data at 3, 4, 5, 5.5, and 6 meV are shifted by 0.6, 1.2, 1.8, 2.4, and 3.2 along the y axis, respectively. The upward arrow in (a) and (b) indicates the direction of increasing energy transfer.  
      }
\label{INS_2}
\end{figure}

\section{Results and discussion}
\label{GaVS_result}

Figure~\ref{Powder_INS_GaVS} presents the powder-averaged INS spectra of GaV$_4$S$_8$ measured using SEQUOIA with $E_i = 18$~meV at $T=5$ and 20~K. In the FM phase at 5 K, the spectra exhibit a single excitation emerging from $Q=0$ with a bandwidth of $\sim6$ meV. At larger wave-vector transfers the scattering decreases in intensity consistent with the cross-section being magnetic in origin. With a higher incident neutron energy of $E_i$ = 40 meV (not shown), we confirm that no additional magnetic scattering is observed beyond that evident in Fig. \ref{Powder_INS_GaVS}(a).  At 20 K (Fig. \ref{Powder_INS_GaVS}(b)),  weak diffuse inelastic scattering is observed. This indicates that the spin-spin correlations survive above $T_N=12.8$~K as expected for a material with geometrical frustration such as \gvs{}.

The INS data collected on the single crystal array at $T=1.7$~K in the FM phase are summarized in Fig. \ref{INS_1}. Reciprocal lattice unit (r.l.u.) indices are denoted in the cubic space group for convenience. Spin waves emanating from the ferromagnetic wave vectors are observed along the (001) and (110) directions as shown in Figs. \ref{INS_1}(a) and (b),  respectively.  The periodicity of the FM spectrum can clearly be seen in the constant-energy slice in the ($HHL$) plane at an energy transfer of 3 meV as shown in Fig. \ref{INS_1}(c). Along both the (001) and (110) directions, the energy of the spin wave mode increases smoothly to $\sim 5.7$~meV at the zone boundary, which is consistent with the bandwidth observed in the powder sample measurement.

To understand the spin excitations in \gvs, we develop an effective Hamiltonian based on molecular $V_4$ spins. Recently Dally et al. have argued that the spin distribution is uniform over the V$_4$ tetrahedra~\cite{Rebecca_2020}. Therefore, for the description of the spin wave dispersion in \gvs, the breathing pyrochlore lattice formed by the V ions can be effectively treated as a face-centered cubic (FCC) lattice of the $V_4$ molecular spins.  %A similar approach has been used to understand the spin-spin interactions in the breathing pyroclores LiGaCr$_4$S$_8$ \cite{pokharel_cluster_2020} and CuInCr$_4$S$_8$ \cite{gao_hierarchical_2021}.
A similar approach has been used to understand the spin-spin interactions in the Cr-breathing pyroclore materials \cite{Ghosh2019,pokharel_cluster_2020, gao_hierarchical_2021}. Unlike the aforementioned Cr-based materials where intra-cluster interactions are evident, the uniform spin distribution over the $V_4$ tetrahedra only modulates the intensities through a magnetic form factor $\int d\bm{r} \exp(-i\bm{kr})\rho(\bm{r})$, where $\rho(\bm{r})$ describes a ferromagnetic spin density that is equally distributed over the 4 V sites, $\bm{k}$ is the wavevector transfer. It should be noted that a recent neutron diffraction experiment on GaV$_4$S$_8$ proposed the spins within the V$_4$ tetrahedra may be tilted by an angle of 39(8)$^\circ$~\cite{holt_investigation_2021}. Such an uncertainty in spin alignment may modify the exact form of the $V_4$ form factor but will not qualitatively change our effective model as long as the major spin components are ferromagnetic.  The coupling strengths between the V$_4$ spins can then be parameterized by comparing the experimental data to the dynamical spin structure factor on a FCC lattice.
Using the linear spin wave theory implemented in the SpinW software package~\cite{Toth_2015}, we consider an effective model with NN exchange and DM interactions:
\begin{equation}
H = \sum\limits_{\braket{ij}}J\bm{S}_i\cdot\bm{S}_j + \sum\limits_{\braket{ij}}\bm{D}\cdot(\bm{S}_i\times \bm{S}_j) \textrm{,}
\label{J_DM_GaVS}
\end{equation}
where $\langle ij\rangle$ denotes the NN bonds with spins $\bm{S}_i$ and $\bm{S}_j$ located at the centers of the V$_4$ tetrahedra. The spin magnitude is assumed to be $S = 1/2$ per V$_4$ tetrahedra, which is consistent with the ordered moment of $\sim$ 0.25 $\mu_B$ per V ion~\cite{Rebecca_2020}. The direction of the DM vector is fixed based on the FCC lattice symmetry. According to the cycloidal pitch and the mean-field analysis of magnetization measurements reported in 
Ref. \cite{Loidl1500916}, we fixed the DM interaction to the reported value of $D$ = 0.13 meV. %Fitting the calculated values to the experimental dispersion along the measured high symmetry directions, using a $\chi^2$-minimization technique, we obtain an optimized fitting parameter of $J = -0.72(3)$~meV. 
To fit the data, we extracted 26 different energy and momentum values from the experimentally observed spin waves along the high symmetry directions (110) and (001). A particle swarm oscillation (pso) fitting method implemented in the SpinW software package is used to fit the data. The spin wave modes within the energy bin size are binned together and considered as one mode in the fit. The fit to the 2D dispersion data is carried out multiple times with more than 50 fitting cycles at each time. The goodness of fit (or R-value) of all those fits is around 2.7. Based on the value of the fitting parameter after all those runs; the value of exchange interaction $J$ and its uncertainty is optimized as -0.72(3) meV. As already noted, the DM interaction is fixed based on the cycloidal pitch and the mean field analysis of magnetization measurements reported in Ref.  \cite{Loidl1500916}. The ratio of $D$ and $J$ in GaV$_4$S$_8$ is around 0.18 which is comparable to many other skyrmion host materials. For example, $D/J$ $\sim$ 0.18 stabilizes the skyrmion state in MnSi \cite{Iwasaki2013, PhysRevB.93.195101}. %The magnetic skyrmions in monolayers of the chromium trihalides $Cr(I,Br)_3$, $Cr(I, Cl)_3$ are characterized by $D/J$ ratios of 0.15 and 0.19 respectively \cite{PhysRevB.101.060404}.
Whereas the nearest neighbor $D/J$ value of 0.39 in the skyrmion host bulk material Cu$_2$OSeO$_3$ \cite{PhysRevLett.109.107203} is larger than its value for GaV$_4$S$_8$. In the skyrmion ground state of a series of Mn$_{1-x}$Fe$_x$Ge ($0<x<1$) compounds \cite{Shibata2013}, the value of $D/J$ varies in the range of (0.02 to 0.6) depending upon the value of x.  %In the skyrmion ground state of PdFe bilayer deposited on Ir(111), the value of $D/J$ vary in the range of (0.05 to 0.2) depending upon the inward Fe layer relaxation \cite {PhysRevB.90.094410}.  

The simulated powder average signal is presented in  Fig.{}\ref{Powder_INS_GaVS}(c) and 2(d). A model obtained by accounting the form factor of V$_4$ tetrahedra, Fig. \ref{Powder_INS_GaVS} (d), reproduces the measured powder spin spectra with better reliability than a model with the form factor of single V$^{3+}$ ion, Fig. \ref{Powder_INS_GaVS} (c), providing further support for the molecular picture for the V$_4$ tetrahedra.  The exchange parameters reproduce spin wave like powder average excitation with zone center $|Q|$ = 0. The calculated energy at each $|Q|$ overlaps with the experimentally measured excitation energy.  As the form factor of V$_4$ tetrahedra changes significantly, our model  indicates that the intensity of excitation spectrum decreases rapidly with the increase in the value of $|Q|$, consistent with the experimental observation. The simulated spin wave spectra, including all domain contributions, for single crystal measurements are shown in the lower panels of Fig. \ref{INS_1}. The energy scale and the features of the calculated excitation spectrum are in accord with the experimental data. 

Including further neighbor interactions does not result in a discernible increase in the quality of the fit with the present data set. Additionally, we note that the observed ferromagnetic ground state originates in part due to anisotropy which stems from the rhombohedral structural distortion at 44 K.  Such anisotropic terms are to too small to be extracted from the data and analysis presented here.

Figures~\ref{INS_2}(a) and (b) present the scattering intensity at fixed energy transfers as a function of L along the (-1 -1 L) and H along the (H H -1) directions. Near the zone boundary, the line-cuts shows broader peak-width along the (H H -1) direction compared to the (-1 -1 L) direction. The calculated constant energy line-cuts at the corresponding energies are over-plotted to the data. The calculated line-cuts successfully captures the peak position and broadness present in the data. The deviation between the model and the data near 5.5 meV is likely due to larger background scattering in this region of the spectrum.

 We now compare the value of exchange coupling determined here with the value determined through other approaches \cite{Wang_2017,Loidl1500916}. Assuming the rhombohedral structure, Zhang et al. \cite{Wang_2017} calculated the NN (interplane) exchange coupling, using first-principles calculations and symmetry analysis. They found values in the range of [$-0.16$, $-1.68$]~meV depending on the value of the effective Coulomb potential. They reported a ratio $D/J$ of $\sim$ 0.16 stabilizes the noncollinear spin structures in GaV$_4$S$_8$, which is consistent with our model.
 The scenario of different coupling strengths due to the rhombohedral distortion is also discussed in Ref.~\cite{Wang_2017}. Mean-field analysis of magnetization measurements in the Ref. \cite{Loidl1500916} showed the NN exchange interaction along the easy axis $\sim$-0.38 meV and an extremely weak magnetic anisotropy with a ratio of exchange couplings parallel and perpendicular to the easy axis $\sim$1.08. The single magnon mode observed in our MACS experiment suggests the couplings strengths among the V$_4$ tetrahedra to be close to uniform even in the rhombohedral phase. As noted previously, the experimental resolution of our measurements is not sufficient to distinguish small anisotropy terms, our value of exchange coupling $J$ is fairly consistent with the value reported through other approaches.

%*****
\section{Conclusions}

We have successfully grown single crystals of GaV$_4$S$_8$ of sufficient size to study the spin dynamics using INS. Our INS experiments reveal well-defined spin wave excitations with a bandwidth of $\sim6$~meV, and an effective FCC lattice model of molecular $V_4$ spins reproduces the INS spectra. In addition to the ferromagnetic couplings of $J$ = -0.72(3) meV between neighboring V$_4$ spins, strong DM interactions with a magnitude of $D\sim0.13$ meV are also confirmed. Our work clarifies the spin couplings in GaV$_4$S$_8$ and will facilitate the further understanding of its rich phase diagram.

\begin{acknowledgments}
We thank C. Batista for useful discussions. This work was supported by the U.S. Department of Energy, Office of Science, Basic Energy Sciences, Materials Sciences and Engineering Division. G.P. acknowledges partial support from the Gordon and Betty Moore Foundations EPiQS Initiative Grant No. GBMF4416. G.P. also acknowledges  support from UC Santa Barbara Quantum foundry through NSF
DMR-1906325.  H.S.A. acknowledges support from the Gordon and Betty Moore Foundation’s EPiQS Initiative Grant No. GBMF9069. This research used resources at the Spallation Neutron Source and the High Flux Isotope Reactor, a Department of Energy (DOE) Office of Science User Facility operated by Oak Ridge National Laboratory (ORNL). Access to MACS was provided by the Center for High Resolution Neutron Scattering, a partnership between the National Institute of Standards and Technology and the National Science Foundation under Agreement No. DMR-1508249.
\end{acknowledgments}

%\bibliographystyle{apsrev4-1}
%\printbibliography
%\bibliography{GaVS.bib}
%\printbibliography

\end{document}